\documentclass[twocolumn]{aa}

\usepackage{graphicx}
\usepackage{multirow} 
\usepackage{threeparttable}
\usepackage{booktabs} 
\usepackage{siunitx} 
\usepackage{amsmath} 
\usepackage{subcaption}
\usepackage{natbib}
\bibpunct{(}{)}{;}{a}{}{,} 
\usepackage[colorlinks=true, linkcolor=blue, citecolor=blue, urlcolor=blue]{hyperref}

\newcommand{\diffp}[2]{\frac{\partial #1}{\partial #2}}

\newcommand{\vb}[1]{\boldsymbol{#1}}

\newcommand{\pdv}[3][]{\frac{\partial^{#1} #2}{\partial #3^{#1}}}

\makeatletter
\let\linenumbers\relax
\makeatother

\begin{document}

   \title{DARWEN: Data-driven Algorithm for Reduction of Wide Exoplanetary Networks}
   \subtitle{An unbiased approach to accurately reducing chemical networks}

   \author{A. Lira-Barria \inst{1}
          \and
          J. N. Harvey \inst{1}
          \and
          T. Konings \inst{2}
          \and 
          R. Baeyens \inst{3}
          \and
          C. Henríquez \inst{4}
          \and  
          L. Decin \inst{2}
          \and
          O. Venot \inst{5}
          \and
          R. Veillet \inst{5}
        }

   \institute{Department of Chemistry, KU Leuven,  Celestijnenlaan 200F, 3001 Leuven, Belgium \\
              \email{arturo.lira@kuleuven.be} \and
              Institute of Astronomy, KU Leuven,  Celestijnenlaan 200D, 3001 Leuven, Belgium
        \and
              Anton Pannekoek Institute for Astronomy, University of Amsterdam, Science Park 904, 1098 XH, Amsterdam, The Netherlands
         \and
              Millennium Institute Foundational Research on Data, Vicuña Mackenna 4860, Macul, Santiago, Chile
         \and 
             Université Paris Cité and Univ Paris Est Creteil, CNRS, LISA, F-75013 Paris, France
             }

   \date{Received August 30, 2024; accepted November 18, 2024}

\abstract
   {Exoplanet atmospheric modeling is advancing toward complex coupled circulation-chemistry models, from chemically diverse 1D models to 3D global circulation models (GCMs). These models are crucial for interpreting observations from facilities like JWST and ELT and understanding exoplanet atmospheres. However, maintaining chemical diversity in 1D models and especially in GCMs is computationally expensive, limiting their complexity. Optimizing the number of reactions and species in the simulated atmosphere can address this tradeoff, but there is a lack of transparent and efficient methods for this optimization in the current exoplanet literature.}
   {We aim to develop a systematic approach for reducing chemical networks in exoplanetary atmospheres, balancing accuracy and computational efficiency. Our method is data-driven, meaning we do not manually add reactions or species. Instead, we test possible reduced chemical networks and select the optimal one based on metrics for accuracy and computational efficiency. Our approach can optimize a network for similar planets simultaneously, can assign weights to prioritize either accuracy or efficiency, and is applicable in the presence of photochemistry.}
   {We propose an approach based on a sensitivity analysis of a typical 1D chemical kinetics model. Principal component analysis was applied to the obtained sensitivities. To achieve a fast and reliable reduction of chemical networks, we utilized a genetic algorithm (GA), a machine-learning optimization method that mimics natural selection to find solutions by evolving a population of candidate solutions.}
   {We present three distinct schemes tailored for different priorities: accuracy, computational efficiency, and adaptability to photochemistry. These schemes demonstrate improved performance and reduced computational costs. Our work represents the first reduction of a chemical network with photochemistry in exoplanet research.}
   {Our GA-based method offers a versatile and efficient approach to reduce chemical networks in exoplanetary atmospheres, enhancing both accuracy and computational efficiency.}

   \keywords{Astrochemistry --
                Methods: numerical --
               Planets and satellites: atmospheres --
               Planets and satellites: composition --
               Planets and satellites: gaseous planets
               }

   \maketitle
%
\section{Introduction}\label{intro}

The field of exoplanet atmospheric modeling is advancing toward more complex coupled circulation-chemistry models, including global circulation models \citep[GCMs;][]{Drummond2020, Ridgway2023, Zamyatina2023}. These models are crucial for accurately interpreting observations from advanced facilities like the \textit{James Webb} Space Telescope (JWST) and the Extremely Large Telescope (ELT). However, the computational expense of GCMs often limits the chemical diversity within these models, necessitating the development of reduced chemical networks \citep{Venot2019, Venot2020}.

Despite their simplicity and the challenge of capturing a planet's entire atmospheric chemistry in a single air column, 1D models have significantly advanced our understanding of hot exoplanets' atmospheric chemistry \citep{Moses2011, Venot2012}. These models reveal that a hot Jupiter's atmosphere has three distinct chemical regimes. In the lower atmosphere, high temperatures and pressures ensure chemical equilibrium. Ascending to the mid-atmosphere, vertical mixing outpaces chemical reactions, leading to a steady state distinct from chemical equilibrium that is invariant with altitude for the most abundant species. This transition from the lower to the middle atmosphere takes place at the quenching level, whose altitude varies depending on the species. Farther up, the highest part of the atmosphere is dominated by photochemistry due to stellar radiation. This framework, although a simplification, provides crucial insights into the dynamic processes governing atmospheric compositions across different height layers \citep{Madhusudhan2016}.

Observations of exoplanet atmospheres, using both ground-based telescopes and space observatories, have detected key molecules such as $\mathrm{H}_2\mathrm{O}$, $\mathrm{CO}_2$, $\mathrm{CO}$, and $\mathrm{CH}_4$ \citep[e.g.,][]{Kreidberg2014, Xue2023, Alderson2023, Esparza-Borges2023, Finnerty2024, Guilluy2019, Bell2023}, alongside less frequently observed molecules like $\mathrm{OH}$, $\mathrm{C}_2\mathrm{H}_2$, $\mathrm{HCN}$, and $\mathrm{NH}_3$ \citep{Nugroho2021, Landman2021, Brogi2023, Giacobbe2021, Cabot2019, Basilicata2024}. All these molecules are crucial for a comprehensive understanding of exoplanet atmospheres. These molecules are particularly significant as they are involved in pivotal thermal and chemical processes and are essential for precise spectral analysis. Moreover, measuring the abundances of these species plays a crucial role in estimating the carbon-to-oxygen (C/O) ratio, an established marker for understanding planetary formation and evolutionary history \citep{Madhusudhan2012}. Recent studies have also begun to explore nitrogen and sulfur ratios, providing additional insights into planetary formation processes \citep{Kama2019, Turrini2021, Ohno2023, Crossfield2023}. Together, these molecules serve as key indicators not only for compositional analysis but also for the complex interplay of chemical dynamics within exoplanet atmospheres. While models ideally aim to be as complete as possible, prioritizing the accurate representation of this list of key species can often suffice for many practical analytical purposes.

Water and carbon dioxide are common in hydrogen-dominated, high-temperature environments \citep[e.g.,][]{Rustamkulov2023}, while methane, ammonia, hydrogen cyanide, and acetylene can indicate disequilibrium processes and the presence of hydrocarbons \citep{Moses2011, Rimmer2019}. The list of key molecules may expand in the future. For instance, the recent recognition of $\mathrm{SO}_2$ \citep{Alderson2023, Dyrek2024, Tsai2023} underscores the importance of considering disequilibrium chemistry, particularly photochemistry, in atmospheric models \citep{Tsai2024, Baeyens2024}. Observations show that key species like $\mathrm{SO}_2$ can drastically deviate from thermochemical equilibrium and significantly impact observables, highlighting the inadequacy of assuming chemical equilibrium when explaining new data. Additionally, the number of studied exoplanet atmospheres is rapidly increasing, enabling the simultaneous detection of several atomic species \citep{Prinoth2023, Mounzer2022}. This expansion has prompted further surveys \citep{Hoeijmakers2019, lira2021} and the first population studies \citep{Welbanks2019, Changeat2022, Edwards2023}. As observational capabilities advance, the need for comprehensive kinetic networks to accurately model these atmospheric compositions in the presence of disequilibrium becomes critical. The growing complexity of models requires efficient and accurate methods for reducing the computational expense of these more diverse chemical models, ensuring they remain effective tools for interpreting the complex dynamics and chemistry of exoplanet atmospheres.

To ensure accurate chemical modeling for exoplanets, \citet{Venot2012, Venot2015, Venot2019, Venot2020} developed comprehensive chemical networks applicable to various contexts, from hot Jupiters to the Solar System's gas giants, and best suited for hot, hydrogen-dominated atmospheres (which means they are not well suited for temperatures below 300 K). These networks were meticulously built and validated through comparison to experimental combustion data \citep{Veillet2024}. In particular, \citet{Venot2020} updated their full chemical scheme (referred to here as V20), which comprises approximately 100 species and 2000 reactions. Using the methodology introduced in \citet{Venot2019}, they reduced it to a more compact form (referred to here as R20), consisting of 44 species and around 600 forward and reverse reactions. This careful reduction significantly lowered computational demands compared to the full model for a set of key molecules while preserving accuracy. The reduction was achieved through a detailed kinetic analysis using ANSYS© Chemkin-Pro Reaction Workbench and incorporating classical techniques from \citet{Lebedev2013}, alongside expert assessment of the impact of removing steps and species in the reduced model.

Building on this previous work, we propose here a new and automated method for generating optimized reduced models that satisfy various criteria such as computational efficiency. We achieved this by using a genetic algorithm (GA) to efficiently reduce chemical networks. The GA simultaneously explores multiple solutions, effectively bypassing local optima. This technique, operating analogously to natural selection, progresses only the most promising solutions based on a predefined fitness function, thus navigating through the most beneficial regions of the solution space without needing derivative calculations \citep{goldberg89}. While GAs have been applied in chemical network reduction \citep{Edwards1998, Elliott2005, Sikalo2014}, including some astrochemical networks \citep{Xu2019, Grassi2012, Grassi2013}, we are not aware of any instances when they have been used in this way for modeling exoplanetary atmospheres. In this work, we use V20 as a reference network to evaluate our reduction method. We note that this network is continually being improved \citep{Veillet2024}, but we do not take the new updates into account here.

The efficacy of this method was tested on hot giant exoplanets with substantial atmospheres, ideal candidates for atmospheric characterization due to their numerous studies and observable features. We specifically chose HD 209458b and HD 189733b as test cases; both planets have been subjects of comprehensive atmospheric composition studies \citep[e.g.,][]{Redfield2008, astudillo2013, sanchez-lopez2019, lira2021, Xue2024, Blain2024}. These studies have examined a variety of atmospheric phenomena, including dynamics, cloud formation \citep{Moses2011, Lines2018}, thermal inversions \citep{Diamond-Lowe2014}, and atmospheric escape \citep{etangs2012, lampon2020}, establishing them as critical benchmarks for validating new chemical reduction techniques.

In this paper we introduce our new GA-based method; it has been designed to be both unbiased and adaptable and is capable of balancing accuracy and computational efficiency across various planetary conditions, including those involving photochemistry. In Sect. \ref{tools} we describe the 1D model we used and the chemical network targeted for reduction. Section \ref{init_reduc} details our initial reduction technique, which involves a sensitivity analysis and a reduction based on principal component analysis (PCA). The core component, our GA-based Data-driven Algorithm for Reduction
of Wide Exoplanetary Networks (DARWEN),  is explained in Sect. \ref{GA}. In Sect. \ref{results} we present three output types tailored to different priorities -- accuracy, computational efficiency, and adaptability to photochemistry -- discussing their respective advantages and limitations. Finally, we conclude with our findings and future directions in Sect. \ref{conclusions}.


\section{Modeling tools}\label{tools}

\subsection{1D chemical kinetics and photochemistry model}\label{model}
We used and modified the 1D chemical kinetics code from \citet{agundez2014}, which takes vertical mixing, molecular diffusion, and photochemistry into account, specifically utilizing its existing option to exclude horizontal mixing. Starting from initial abundances at chemical equilibrium at each pressure level, the model integrates the system of ordinary differential equations (ODEs) as a function of time. These ODEs are solved using the DLSODES \citep{Hindmarsh1983} Fortran library. 
This process continues until a steady state is reached, capturing the chemical evolution of the atmosphere over time. This steady state differs from chemical equilibrium because, while chemical reactions and transport processes balance each other out, the movement of molecules to different parts of the atmosphere causes the chemical composition to deviate from thermochemical equilibrium. The equation describing the formation and destruction of species is given by Eq.~\ref{eq:continuity_f}:

\begin{equation} \label{eq:continuity_f}
   \diffp{f_i}{t} = \frac{P_i}{n} - f_i L_i - \frac{1}{nr^2} \pdv{ \left(r^2 \Phi_i \right) }{r}
.\end{equation}In this equation, $f_i = n_i/n$ represents the molar fraction, also referred to as the abundance, of each species $i$, where $n_i$ is the species' number density, and $n$ is the total number density of all particles. The $f_i$ depends both on time ($t$) and the distance from the center of the planet ($r$), or equivalently the altitude ($h=r-r_0$); hence, a specific molar fraction is referred to as $f^t_{ih}$. The terms $P_i$ and $L_i$ denote the production and loss rates of species $i$, respectively, incorporating both chemical reactions and photochemical processes. The model also accounts for the vertical transport flux $\Phi_i$ of each species, which may be either positive (indicating upward movement) or negative (indicating downward movement), driven by atmospheric dynamics including eddy and molecular diffusion. This dual approach captures both larger-scale atmospheric movements driven by eddy diffusion and smaller-scale, molecular-level movements within the atmosphere \citep[for details, see][]{agundez2014, bauer1973}.

\subsection{Chemical network and other data}\label{inputs}

To properly model an atmosphere, the code needs several inputs. As the physical state of the atmosphere, we used the pressure-temperature and eddy diffusion profiles from the model grid of \citet{Baeyens2021, Baeyens2022}. Specifically, we chose a model with ($T_{\mathrm{eff}}$ = 1200 K, $g$ = 1000 cm s$^{-2}$, K5-star host) and a model with ($T_{\mathrm{eff}}$ 1400 K, $g$ = 1000 cm s$^{-2}$, G5-star host), as these parameters are representative of the hot Jupiters HD 189733 b and HD 209458 b, respectively. For initial elemental abundances, we assumed solar metallicity. The temperature profiles were derived from a 3D GCM \citep{Carone2020} at the substellar point, with an isothermal upper atmosphere extension to a minimum pressure of $10^{-8}$ bar. Eddy diffusion coefficient profiles were self-consistently computed from the wind speeds \citep{Baeyens2021}. 

Photochemical dissociations were computed based on the stellar UV spectra of a K5- and a G5-type star. For the former, we used the spectrum of HD 85512 from the MUSCLES survey \citep{France2016, Youngblood2016, Loyd2016}, while for the latter, we used a solar spectrum \citep{Claire2012}, following the approach in \citet{Baeyens2021}. The photochemical cross sections and quantum yields used are based on laboratory data and ab initio calculations \citep{Venot2012, Venot2020, Hebrard2013, DOBRIJEVIC2014}.

Finally, and key for this work, the code requires a chemical network to compute the production and loss rates of abundances at each time iteration. For this study, we used the comprehensive full V20  model from \citet{Venot2020}, which we aimed to reduce. This network comprises 108 species (including C, H, N, and O atoms) with 1906 forward and reverse reactions. Appendix ~\ref{app:reacs} explains how these reactions are treated in the code.

\subsection{Convergence of the model}\label{convergence}

Our convergence values were measured using what we call the maximum relative deviation ($\underset{\text{max}}\Delta$):

\begin{equation}\label{eq:MRD}
    \underset{\text{max}}{\Delta}^{(t_n, t_{n+1})}_{(f_{ih}^{t_n} > f_{min})} = \max_{i, h} \left( \frac{|f_{ih}^{t_{n+1}} - f_{ih}^{t_n}|}{f_{ih}^{t_n}} \right)
.\end{equation}

In this 1D model, $\underset{\text{max}}{\Delta}$ quantifies discrepancies between the abundances of species $i$ across atmospheric layers $h$ at two sequential time points, $f_{ih}^{t_n}$ and $f_{ih}^{t_{n+1}}$. For this purpose, we considered only significant abundances (i.e., those satisfying $f_{ih}^{t_n} > f_{min}=10^{-20}$) to ensure focus on relevant chemical actors. The time points in the model are exponentially spaced after an initial integration time of one day, progressing to 10 days, 100 days, and so on. Chemical steady-state is considered reached once the $\underset{\text{max}}{\Delta}$ of Eq.~\ref{eq:MRD} drops below the threshold of $10^{-4}$ for two successive integration periods. \footnote{The ODE solver's precision is set with a relative tolerance of $10^{-4}$, balancing computational efficiency and accuracy to match the target threshold. The absolute tolerance is set to an extremely small value ($10^{-99}$), ensuring it has no impact on the solution.}

Although a stricter convergence to steady-state could be ideal in principle, the threshold used here is sufficient for conducting the sensitivity analysis and subsequent model reduction. Additionally, we observe that reaching a steady state was difficult when including the lowest layers of the atmosphere, with pressures higher than $10$ bar or, in cases where photochemical processes are included in the model, when modeling high altitudes with pressures below $10^{-5}$ bar. As these pressure ranges do not qualitatively impact on the model outcomes, we therefore excluded them during the optimization process. More specifically, when developing models without photochemistry, we selected a range between $10$ and $10^{-3}$ bar, encompassing the quench points. With photochemistry, we extended the range to $10^{-5}$ bar to include higher altitudes where photochemical processes are significant, while still achieving convergence to our chosen noise threshold.

\section{Initial reduction technique}\label{init_reduc}

Our primary method is a GA, which requires an initial starting point to approximate the target area for finding a more efficient and accurate chemical scheme. To establish this starting point, we performed a sensitivity analysis (detailed in Sect.~\ref{sensis})  to assess how variations in each reaction impact steady-state abundances. Based on these data, we ranked reactions using a PCA-based reduction process (detailed in Sect.~\ref{pca}), which helped us identify the most influential reactions for the subsequent optimization.

\subsection{Sensitivity analysis}\label{sensis}
    A straightforward approach to ranking the importance of parameters in a model, such as reactions, is to perform a sensitivity analysis, either global (varying multiple parameters simultaneously) or local (varying one parameter at a time). While global sensitivity analyses are widely used to reduce models \citep{Saltelli2005, Dobrijevic2010, Perumal2013, Nurislamova2017, Till2019}, it is well known that local sensitivity analysis is insufficient for ranking parameters or reducing models unless the model is linear, as it neglects intrinsic uncertainties. In this study we employed a local sensitivity analysis as an initial approach due to its simplicity and low computational cost; this provided a foundation for further refinement in subsequent steps. 
  
    For this analysis, we evaluated how slight variations in each rate constant, $k_j$, affect species abundances across different atmospheric layers. This process involves iteratively altering one rate constant at a time and observing the resulting changes in abundance of every species at each layer. We quantified these changes using sensitivity matrices, $\vb{S}_j$, for each chemical reaction, $j$, in the model, whose elements measure the dependence of the concentration of a given species at a given altitude on the chosen rate constant. 
  
  The sensitivity coefficients, or elements of the matrix, were calculated based on a finite differences approximation to the derivative of the abundance $f_{ih}$ of species $i$ at altitude $h$, with respect to $k_j$. These differences were computed by re-propagating the model to steady-state with a modified $k_j$, obtained by multiplying it by a factor of $1.1$, which is equivalent to the initial $k_j$ plus $10 \%$ of its value. This factor is large enough to induce a measurable change but small enough to be considered a finite difference.
  
  We note that the finite-step perturbations of the rate constants are not intended here as an approach for uncertainty quantification. In fact, there will be different degrees of uncertainty for each reaction, and the impact of this rate constant inaccuracy is a separate issue that we do not tackle here. Instead, the 10$\%$ differential is used as the basis for ranking the importance of reactions, assuming that reactions causing larger changes are likely to be more significant for network reduction. Additionally, we normalized these sensitivity coefficients with respect to the reference value, $f_{ih}^0$, to ensure that both abundant and less abundant species are treated equitably. Each element of the sensitivity matrix $\vb{S}_j$ is thus given by

\begin{equation}\label{eq:sensi}
    S_{(i,h)}^j = \frac{1}{{ f_{ih}^0}}\diffp{f_{ih}}{k_j} \sim  \frac{\Delta f_{ih}} {f_{ih}^0\Delta k_j} 
    = \frac{f_{ih}^*-f_{ih}^0}{f_{ih}^0 \left(k_j^* - k_j^0 \right)} 
.\end{equation}

\noindent Here, $f_{ih}^0$ is the pre-variation steady-state abundance of species $i$ at each atmospheric layer $h$, while $f_{ih}^*$ represents the post-variation abundance upon reaching a new steady state. Each rate constant for the $\sim$1000 forward reactions present in V20 is varied, and this results in a $\vb{S}_j$ matrix that depends on the 107 species\footnote{We did not take helium into account in this sensitivity analysis.} and 100 height layers for every one of them. If a reaction has an associated reverse reaction, its rate constant is changed by the same factor to preserve thermodynamic consistency (for details, see Appendix ~\ref{app:reacs}).

\subsection{Principal component analysis}\label{pca}

 After obtaining sensitivities for a given planet, we applied the PCA reduction method, as described by \citet{Lebedev2013}. Special emphasis was placed on including key molecules already observed in exoplanetary atmospheres ($\mathrm{H}_2\mathrm{O}, \mathrm{CO}_2, \mathrm{CO}, \mathrm{HCN}, \mathrm{C}_2\mathrm{H}_2, \mathrm{OH}, \mathrm{CH}_4,$ and $\mathrm{NH}_3$) as they are the species we refer to throughout the text as the ``key species'' for evaluating our model's performance. The steps of this method are the following:

\begin{enumerate}

    \item We generated a correlation matrix based on the sensitivity coefficients to evaluate the impact of each reaction. This was done by using the overall sensitivity matrix, denoted as $\mathcal{S}$. This matrix is a combination of individual matrices, $\vb{S}_j$, each calculated according to Eq.~ \ref{eq:sensi}. For any specific reaction $j$, we extracted the elements corresponding to the selected group of key molecules into a single row. These elements are  indexed by a composite column label $(i,h)$. Consequently, the calculation of $\mathcal{S}\mathcal{S}^T$ associates each row with the relative importance of its corresponding reaction.
    
    \item Eigenvalues and eigenvectors were calculated from this matrix, which could then be used to quantify the contribution to the model made by each reaction. The eigenvalues were utilized to filter the contribution of eigenvectors, isolating those with significant impact on chemical dynamics. For the filtered eigenvalues, $\lambda_i$, the corresponding eigenvectors, $u_i$, were projected into the reaction space, $r_j$, with the component of reaction $j$ in the new space given by $r_j^i = (u_i \cdot r_j)/||u_i||$.

    \item We adjusted two thresholds in the PCA-reduced scheme specifically to achieve our target number of reactions and molecules. First, we selected eigenvalues, $\lambda_i$, that cumulatively represent $95\%$ of the overall significance, which typically involves two or three principal components. Second, we set a predefined threshold for the reaction space components, $r_j^i$, based on the magnitude of sensitivities. By fine-tuning these thresholds, we intentionally limited the scheme to around 100 forward reactions, ensuring that we never included more than 50 species.
     
    \item A species was selected for the reduced scheme only if it was involved in at least one of the qualifying reactions. Consequently, the selection of species and reactions for the final PCA scheme was constructed such that the already mentioned key molecules are present. From the 100 forward reactions mentioned in the previous step, the scheme includes approximately 40 species. 
    
\end{enumerate}

We discovered that while our schemes obtained by sensitivity analysis and subsequent PCA-based reduction proved remarkably effective, they yielded lower accuracy than R20, by about one order of magnitude. This discrepancy, we believe, stems from our sensitivity analysis being overly tuned to identify reactions with low reaction rates (see Appendix ~\ref{app:reacs}). This approach leaves out crucial fast reactions, which in the full model precede or follow the most sensitive reactions we obtained, thereby yielding an incomplete network.  However, this network serves as a suitable starting point for our core method.

\section{A genetic algorithm for chemical network optimization}\label{GA}

\begin{figure}[t]
   \centering

   \includegraphics[width=8cm]{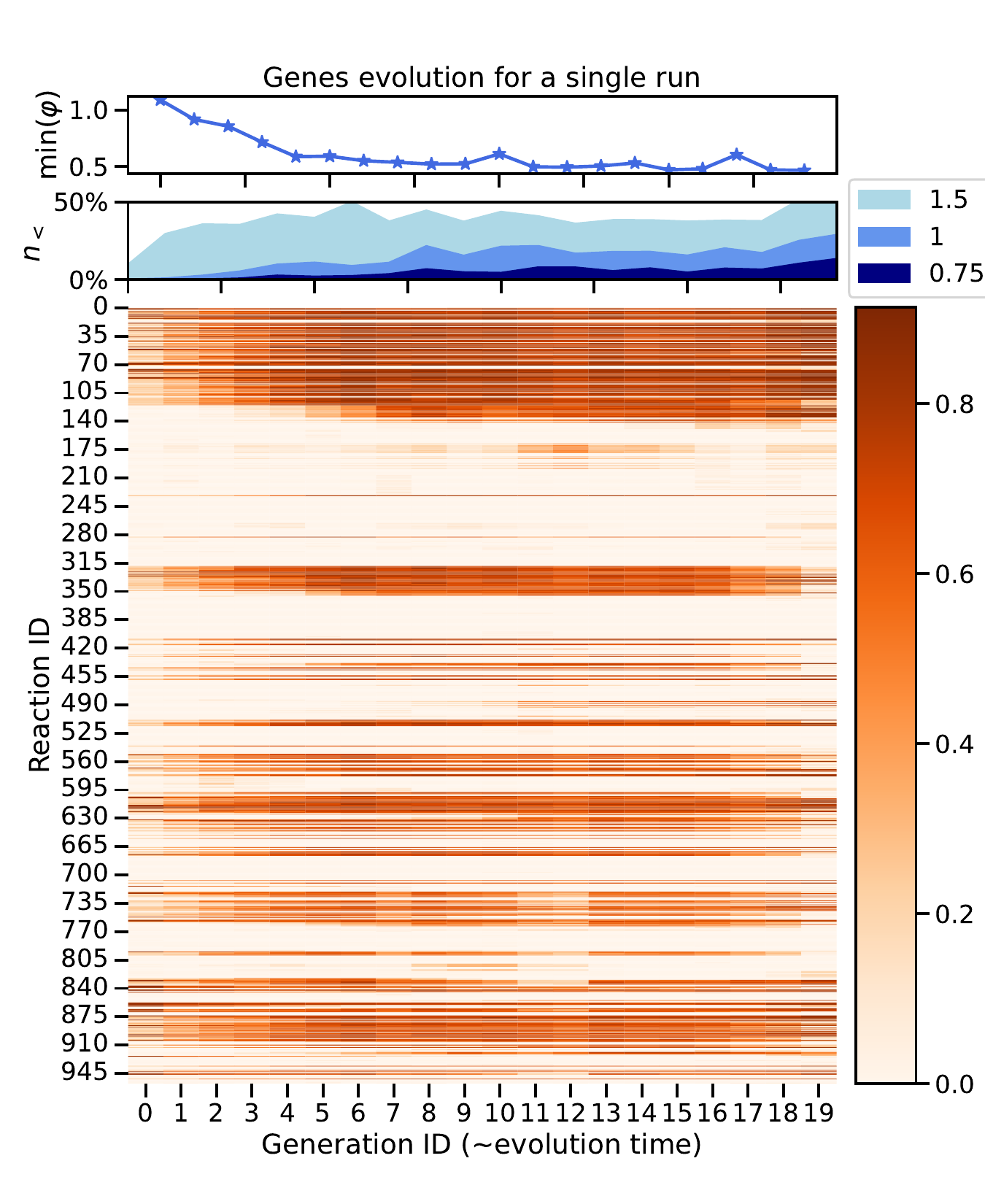}
   \caption{{Example of DARWEN's ``evolution'' during simultaneous optimization of two planets.} \textit{Top:} Minimum of the loss function, $\varphi$, tracked across all schemes (or individuals) in each generation, showing the progress of optimization. \textit{Middle:} Proportion of individuals with a loss function below thresholds of 1.5, 1, and 0.75 per generation, providing insights into the population's overall performance distribution. This visual indicates how many schemes consistently achieve high fitness levels. \textit{Bottom:} Heatmap visualizing the average value of each gene across all individuals per generation. If all schemes possess a value of 1 for a specific gene, the heatmap is orange, whereas for 0 it appears white. This helps us evaluate the diversity within the population and how closely the schemes align with the PCA-based initial model. It also allows us to identify which types of reactions are most prevalent during the optimization process.}

              \label{fig:genes}
\end{figure}

\begin{table*}[t]
    \centering
    \begin{threeparttable}
    \caption{Results for the two planets tested in this work, for each type of scheme.}\label{tab:losses}
    \begin{tabular}{
        l
        S[table-format=3.0] 
        S[table-format=4.0] 
        S[table-format=1.2]
        S[table-format=2.2]
        S[table-format=2.2]
        S[table-format=3.0]
        S[table-format=1.2]
        S[table-format=2.2]
        S[table-format=2.2]
        S[table-format=3.0]
    }
    \toprule
    \toprule
    \multicolumn{1}{l}{Scheme type} & \multicolumn{1}{c}{$n_{\mathrm{mlc}}$} & \multicolumn{1}{c}{$n_{\mathrm{reacs}}$} & \multicolumn{4}{c}{HD 209458b} & \multicolumn{4}{c}{HD 189733b} \\
    \cmidrule(lr){4-7} \cmidrule(lr){8-11}
    & & & 
    {$\underset{\text{max}}{\Delta}_{\mathrm{\{key\}}}^{(\circ, \bullet)}$} & 
    {$\underset{\text{max}}{\Delta}_{\mathrm{\{major\}}}^{(\circ, \bullet)}$} & 
    {$\underset{\text{max}}{\Delta}_{\mathrm{\{all\}}}^{(\circ, \bullet)}$} & 
    {time [s]} & 
    {$\underset{\text{max}}{\Delta}_{\mathrm{\{key\}}}^{(\circ, \bullet)}$} & 
    {$\underset{\text{max}}{\Delta}_{\mathrm{\{major\}}}^{(\circ, \bullet)}$} & 
    {$\underset{\text{max}}{\Delta}_{\mathrm{\{all\}}}^{(\circ, \bullet)}$} & 
    {time [s]} \\
    \midrule
    V20          & 107 & 1904 & {---} & {---} & {---}    & 500  & {---} & {---} & {---}    & 650  \\
    R20          & 44  & 582  & 0.07  & 0.97  & 0.97     & 26   & 0.06  & 0.98  & 0.98     & 33   \\
    Validation   & 47  & 576  & 0.06  & 0.56  & 1.44     & 25   & 0.06  & 0.58  & 12.45    & 31   \\
    Low-cost     & 32  & 298  & 0.28  & 0.29  & 0.29     & 10   & 0.33  & 0.69  & 0.69     & 13   \\
    \midrule
    V20*         & 107 & 1956 & {---} & {---} & {---}    & 900  & {---} & {---} & {---}    & 950  \\
    Photoscheme    & 48  & 756  & 0.16  & 4.39  & 5.12     & 33   & 0.18  & 4.74  & 6.16     & 43   \\
    \bottomrule
    \end{tabular}
    
    \begin{tablenotes} 
    \small
    \item V20, R20 and V20* (full model by \citet{Venot2020} plus the corresponding photo reactions) are shown here for reference. $n_{\text{mlc}}$ and $n_{\text{reacs}}$ are the number of molecules and forward plus reverse reactions for each scheme. All schemes discrepancies are  between the corresponding full model ($\bullet$) and the reduced scheme in question ($\circ$). Time to reach steady state, according to our criteria (see Sect.~\ref{convergence}), is measured via Terminal on a MacBook Pro 2022 with 16 GB of RAM and an M2 chip. 
    \end{tablenotes}

\end{threeparttable}
\end{table*}

A GA is a machine learning optimization technique inspired by natural selection and genetics in biological systems. In a GA, each candidate solution is represented as a string of genes; specifically, in our work, each gene indicates whether a chemical reaction is included (1) or excluded (0) in a candidate chemical scheme.\footnote{Alternatively, each string can be seen as a list of reaction IDs included in the model.} GAs operate by iteratively selecting the fittest individuals in a population according to a loss function. At each iteration (or generation), each selected individual is used to generate multiple new individuals through random variation using the operators described below. Our initial population starts with a single progenitor derived from our PCA-reduced chemical scheme (detailed in Sect.~\ref{pca}).

Genetic algorithms are particularly useful for solving complex optimization problems where the solution space is vast and not easily navigable using traditional gradient-based methods \citep[see][for a comprehensive description]{goldberg89}. According to the building blocks hypothesis \citep{Holland1992}, the effectiveness of GAs arises from their ability to implicitly evaluate approximately $n_{pop}^3$ highly fit gene combinations, even though only $n_{pop}$ individuals are explicitly tested in each generation. This extensive evaluation occurs without additional computational or memory resources. However, a significant risk is premature convergence, where the population lacks diversity among gene combinations, leading to repetitive testing of limited solutions. This issue, described in \citet{Edwards1998}, can be mitigated by maintaining sufficient genetic variation in each generation, thus ensuring continuous exploration of the solution space.

The core of our new reduction method is DARWEN, a GA designed to handle a vast space of $2^{N_{reacs}} \sim 2^{1000}$ possible combinations arising from the number of genes when trying to reduce the network described in Sect. ~\ref{inputs}. Although GAs are known to struggle in extremely large search spaces like $10^{1000}$\citep{Elliott2004}, the space we are dealing with, while large, remains within a manageable range for our GA.

In our implementation, DARWEN adds and removes forward and reverse reactions together, which is why $N_{reacs}$ is 958 instead of 1906 — some reactions lack a reverse counterpart (see Table ~\ref{tab:glossary} for all parameter values). Importantly, DARWEN does not test every possible reaction combination because schemes that fail to conserve mass within our 1D model are disqualified early in the process. Additionally, by limiting the addition of reactions to those associated with the initial species and a few additional molecules, we reduced the potential solution space to fewer than $2^{350}$ possibilities (the typical pool of available reactions is around 350).

At each generation, we selected $n_{\text{fit}}$ individuals as progenitors based on their fitness values calculated using the loss function $\varphi$. Every $n_{\text{elit}}$ generations, if the overall performance declines, we reintroduced progenitors from earlier generations to preserve valuable genetic information and enhance optimization robustness. Each progenitor has a probability of undergoing crossover ($p_c$), where segments of genes are exchanged with another progenitor, introducing variability and potentially beneficial combinations. There is also a probability of self-mutation ($p_m$), where genes are randomly flipped to include or exclude reactions.

To each individual, we applied a mutation rate, $\rho_m$, and a killing rate, $\rho_{\dagger}$, to add or remove reactions, focusing on those associated with the initial species and a limited number of additional molecules. This targeted approach effectively reduced the solution space. To explore a broader solution space and address reactions that are consistently avoided, we introduced ``curious'' reactions,  $n_{\text{cur}}$. These under-explored reactions helped ensure a broader search of the solution space. More details on these processes and the parameter values used are in Appendix~\ref{app:GA}.

After an individual is created and meets the constraints on the number of molecules, we computed its loss function, $\varphi,$ using the model detailed in Sect.~\ref{model}. This loss function is the core of DARWEN and employs  $\underset{\text{max}}{\Delta}$ (Eq.~\ref{eq:MRD}) to evaluate discrepancies between the reduced ($\circ$) and full ($\bullet$) chemical schemes. It focuses on the key molecules mentioned earlier (see Sect.~\ref{pca}), as well as additional ``major'' species defined at each altitude as those with an abundance above a threshold ($f_{ih}^\bullet > f_{\text{min}} = 10^{-20}$). This means that a molecule can be considered a major species at certain altitude layers but not at others, depending on its abundance. The loss function integrates those discrepancies to balance chemical accuracy against computational cost, which is roughly proportional to the number of molecules in each scheme ($n_{\text{mlc}}$).
So if we consider
\begin{equation}\label{eq:abbreviations}
\begin{split}
\underset{\text{max}}{\Delta}\text{\{key\}} &= \max_{\{\text{planets}\}} \left( \underset{\text{max}}{\Delta}_{\{\text{key}\}}^{(\circ,\bullet)} \right) \\ 
\underset{\text{max}}{\Delta}{\text{\{maj\}}} &= \max_{\{\text{planets}\}} \left( \underset{\text{max}}{\Delta}_{\left(f_{ih}^\bullet > f_{\text{min}}\right)}^{(\circ,\bullet)} \right)
\end{split}
,\end{equation}the loss function, $\varphi,$ becomes
\begin{equation}\label{eq:loss}
\varphi = \underset{\text{max}}{\Delta}\text{\{key\}} + w_0 ~ n_{\text{mlc}} + w_1 \log \left( \underset{\text{max}}{\Delta}{\text{\{maj\}}} \right)
.\end{equation}

Weighting factors $w_0=0.1$ and $w_1=1$ are strategically set, meaning they are chosen based on careful consideration to balance the impact of species count and major species fitness within the loss function. These values were determined by aiming for desired results in the schemes. Specifically, if $w_0$ is too large, it can outweigh the other factors, making the model less effective. The inclusion of $w_1$ ensures that the model does not over-fit to just the key molecules. These factors remain adjustable to facilitate the development of faster models with relaxed accuracy requirements. While DARWEN effectively reduces the number of reactions for individual planets, this method aims to create a more general chemical scheme applicable to a group of planets with sufficiently similar chemistry. The algorithm manages multiple atmospheres by choosing the highest $\underset{\text{max}}{\Delta}$ among the planets being optimized for each chemical scheme at every generation. This approach ensures that when a planet with sufficiently similar characteristics is encountered, the existing reduced model can be applied directly, potentially reducing the need to rerun DARWEN.

As stated in Eq.\ ~\ref{eq:loss}, to optimize models for multiple planets simultaneously, we took the maximum value of each scheme metric when tested in the atmospheric models of different exoplanets. Our initial scheme is always based on a PCA-reduced chemical network. Initially, we attempted to combine the PCA-reduced schemes of our two test-case planets, HD 209458b and HD 189733b. However, this approach proved inefficient, especially when integrating photochemistry, as it introduced chemical inconsistencies. These issues arose from combining reactions from the different atmospheric environments of the planets. We found it more effective to use a PCA-reduced chemical scheme from just one planet. This approach avoided the aforementioned issues and provided a sufficient starting point, even when dealing with models without photochemistry.

\section{Three types of reduced chemical schemes}\label{results}

\begin{figure*}[ht]
    \centering
    \begin{subfigure}[b]{0.49\textwidth}
        \centering
        \includegraphics[width=\textwidth]{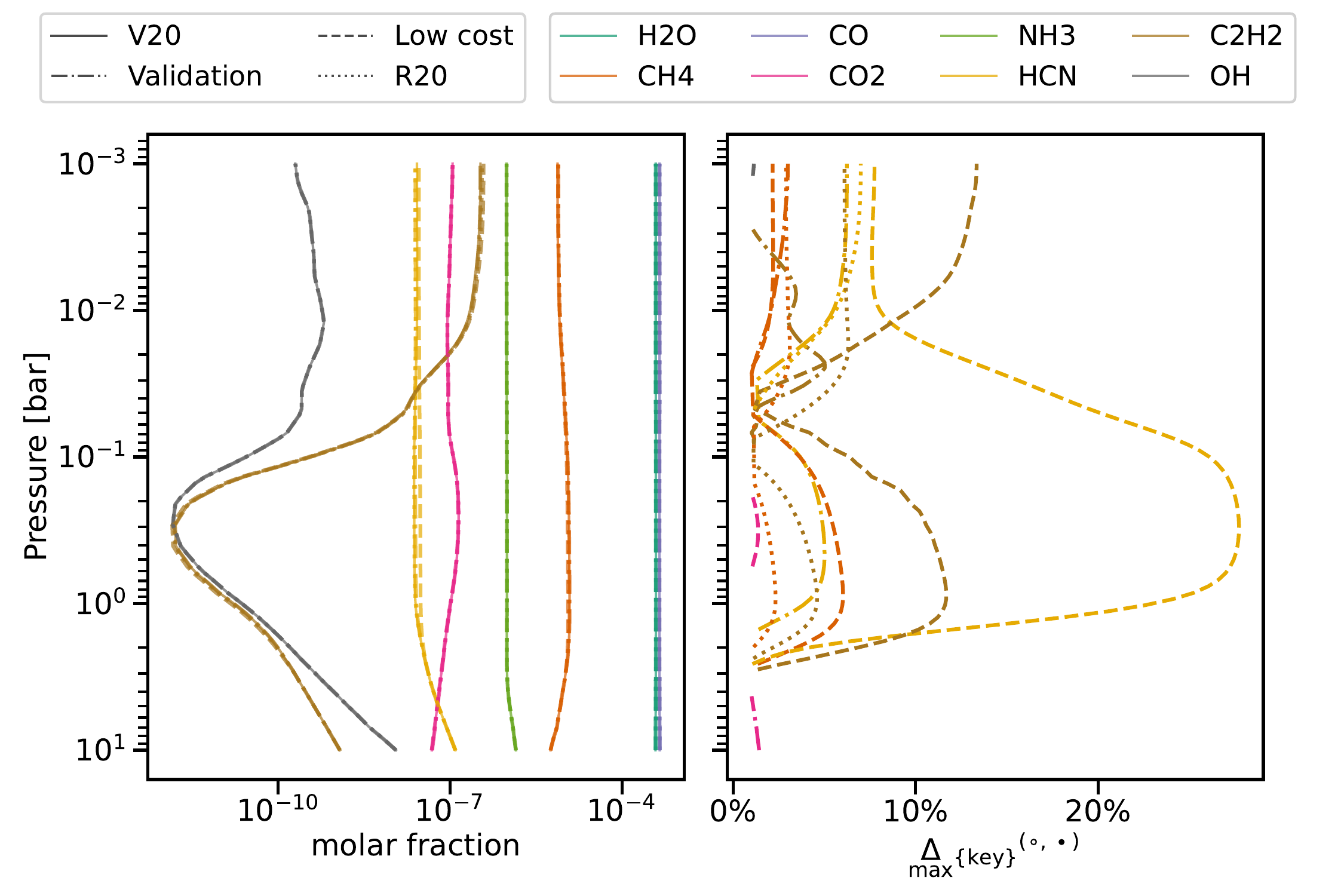}
        \caption{}
    \end{subfigure}
    \hfill
    \begin{subfigure}[b]{0.49\textwidth}
        \centering
        \includegraphics[width=\textwidth]{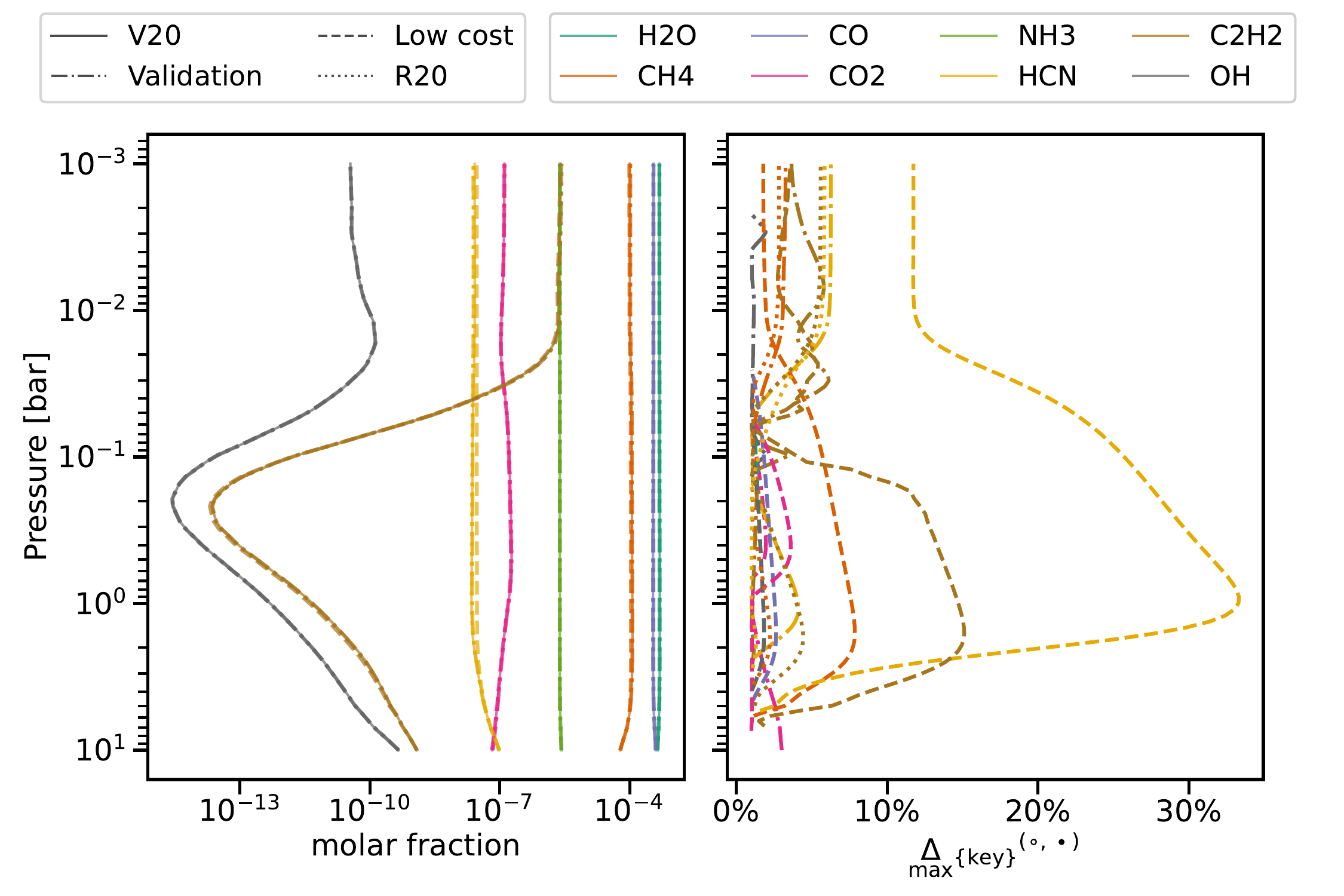}
        \caption{}
    \end{subfigure}
    
    \caption{{Validation and low-cost schemes' performance on key and major molecules.} The plots illustrate the performance of our method on key molecules for HD 209458b (a) and HD 189733b (b). \textit{Left}: Molar fractions of key molecules as a function of pressure in the atmosphere. The full V20 model is represented by a solid line, DARWEN's validation scheme by a dashed-dotted line, low-cost schemes by a dashed line, and the reduced scheme (R20) by a dotted line. \textit{Right}: Maximal percentage changes in the molar fractions of the key molecules compared to the full model. The full model's line is not shown because it serves as the reference model. Lines close to zero indicate that the predictions are nearly identical to those of the full model for that molecule. We only show changes bigger than 1$\%$.}
    \label{fig:key}
\end{figure*}

We applied our method to generate new reduced versions of the V20 network for HD 209458b and HD 189733b's atmospheres. We employed three approaches: schemes specifically aimed at enhancing key molecule accuracy as in previous reduced chemical schemes (we refer to these as validation schemes); schemes that emphasize computational efficiency and thereby restrict the maximum number of species to 40 (low-cost schemes); and schemes that integrate photochemical processes, which is a new development in this field, and should provide optimal accuracy (``photoschemes'').

To ensure the optimal evolution of DARWEN's algorithm, we experimented with several parameters, detailed in Table \ref{tab:glossary}. For the validation and low-cost schemes, we selected a pressure range from 10 to $10^{-3}$ bar, focusing on this interval for three reasons: it captures the entire chemically significant range when photochemistry is not considered, it aligns with the pressures probed by infrared observations, and it leads to efficient convergence of the \citet{agundez2014} model and, hence, rapid optimization. In contrast, for models involving photochemistry, we extended the range to 10 to $10^{-5}$ bar, which is more computationally demanding but essential for capturing the higher altitude regions where photochemical behavior is relevant. Additionally, we note that the inclusion of photochemistry requires either separate optimizations to account for the unique photolytic dynamics of each atmosphere or a less accurate reduced scheme. This is due to the increased computational expense of photochemical processes, which slows and complicates DARWEN's reduction process.

\subsection{Validation scheme}\label{valid}

We allowed DARWEN to evolve over 20 generations, transitioning from a PCA-reduced scheme to an optimal ``validation'' scheme, aiming for precision comparable to V20 for our specific planets. To obtain optimal schemes, we employed a strategic bottleneck approach: after approximately ten generations, we forced the best-performing scheme to become the initial scheme for a subsequent run of DARWEN (i.e., at generation $\sim$ 10, we set $n_{\text{fit}}=1$). An example of this process is shown in Fig. ~\ref{fig:genes}.
Following this optimization, we identified around two dozen candidate schemes to validate our procedure. The final choice was made based on the accuracy of the broader set of the ``major'' (i.e.,\ the most abundant) species. While R20 has 582 reactions (30 $\%$ of the total number of reactions in the full model, and all the reactions available for that subset of 44 species), our validation scheme has 576 reactions (again 30 $\%$ of the total number of reactions in the full model, but 85 $\%$ of reactions available for those 47 species). As detailed in Table~\ref{tab:losses} the $\underset{\text{max}}{\Delta}$ for key molecules in the atmospheres of the two tested planets (0.05-0.06) is comparable to the $\underset{\text{max}}{\Delta}$ in the R20 scheme (0.07-0.06). Taking into account that an ideal case would have the full model abundances identical to the reduced model, resulting in a null $\underset{\text{max}}{\Delta}$, a lower $\underset{\text{max}}{\Delta}$ indicates a more accurate network reduction. This demonstrates DARWEN's capability to generate reduced chemical schemes, preserving accuracy with an error of less than 6$\%$ on key species.

Notably, the $\underset{\text{max}}{\Delta}$ for major species in our model ($\underset{\text{max}}{\Delta}{\text{\{maj\}}}$) was approximately 0.6, compared to around 1 in R20. Since major species generally outnumber key species, their abundances tend to show less precise agreement with those in the full model. While R20 maintained a similar level of discrepancies between all species and major species, we intentionally chose schemes that sacrificed accuracy for minor species to better preserve the accuracy of major species. This balance choice reflects our priorities, as we are generally less interested in minor species regarding their spectral features and overall impact on key species. By adjusting $w_1$ in our loss function $\varphi$, we can control this balance; increasing $w_1$ elevates the relative importance of $\underset{\text{max}}{\Delta}{\text{\{maj\}}}$, leading DARWEN to prefer schemes that are more precise for major species (for more details on important discrepancies on major species, see Appendix~\ref{app:major}).

Furthermore, our selected scheme includes more molecules but fewer reactions. While R20 is $\sim$ 19 times faster than V20, our selected scheme was $\sim$ 20 times faster than V20 (see Table ~\ref{tab:losses}). This efficiency, combined with improved accuracy for chemically significant species, demonstrates the effectiveness of our approach in optimizing chemical network reduction.

\subsection{Low-cost schemes}

We have confirmed that DARWEN is highly efficient at reducing schemes to a level of accuracy similar to R20. We then decided to explore how much we could reduce the number of molecules while maintaining this accuracy. Using our validation scheme as a starting point, we developed the ``low-cost'' scheme. As shown in Table~\ref{tab:losses}, this scheme runs 2.5 times faster than R20 while maintaining comparable accuracy. The largest discrepancy with respect to V20, as shown in Fig.~\ref{fig:key}, occurs for $\mathrm{HCN}$, which has an $\underset{\text{max}}{\Delta}$ of 0.27 for HD 209458b and 0.33 for HD 189733b, though the shape of the abundance profiles as a function of altitude remains well described.

Achieving the presented results required approximately 30 generations, starting from our optimal validation scheme, to balance accuracy for key molecules while minimizing the number of molecules and maintaining overall species accuracy. This published scheme has the lowest number of molecules while meeting our accuracy requirements, showcasing DARWEN's capability to efficiently optimize chemical network reduction, balancing precision and computational cost effectively.

\begin{figure*}[ht]
    \centering
    \begin{subfigure}[b]{0.49\textwidth}
        \centering
        \includegraphics[width=\textwidth]{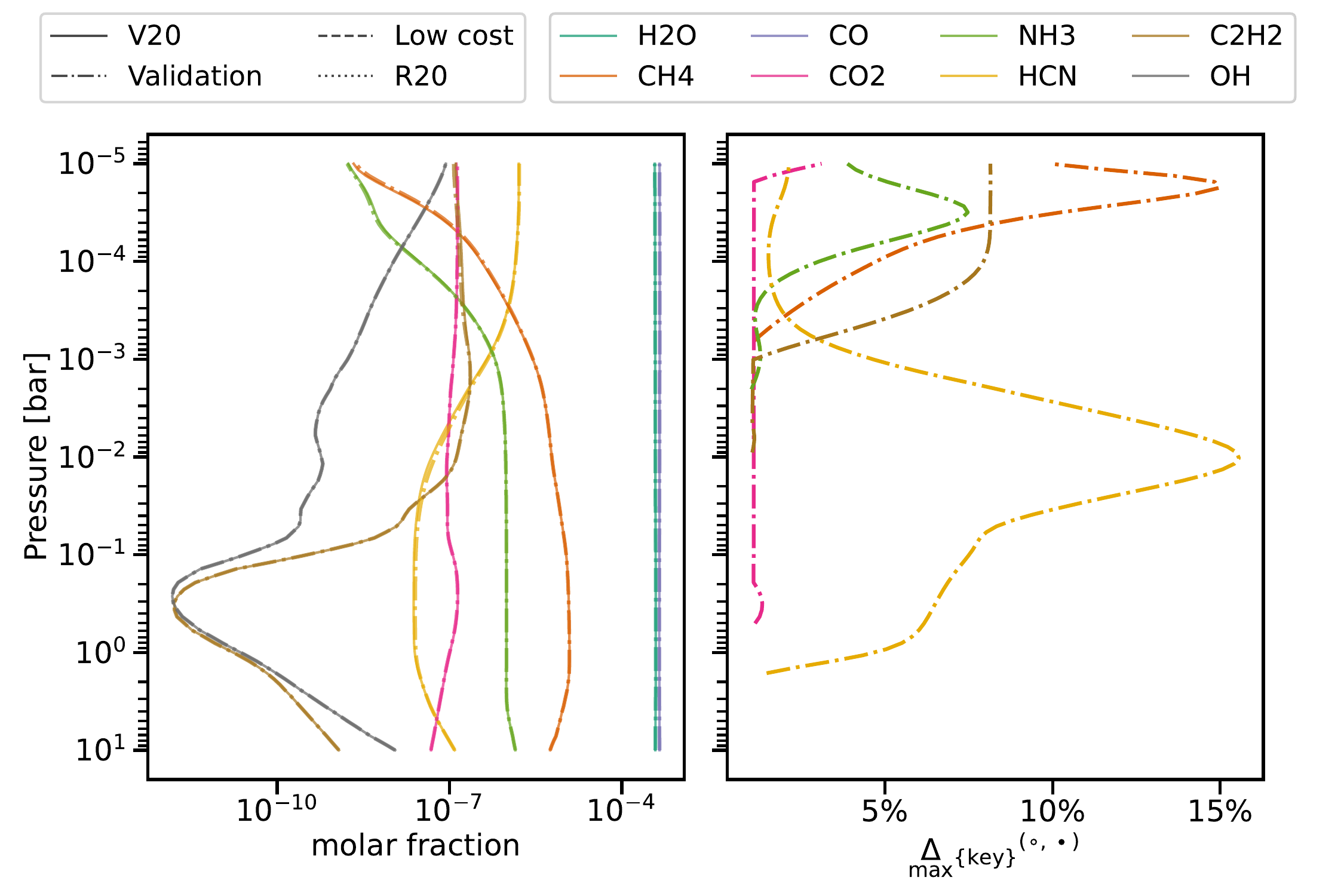}
        \caption{}
    \end{subfigure}
    \hfill
    \begin{subfigure}[b]{0.49\textwidth}
        \centering
        \includegraphics[width=\textwidth]{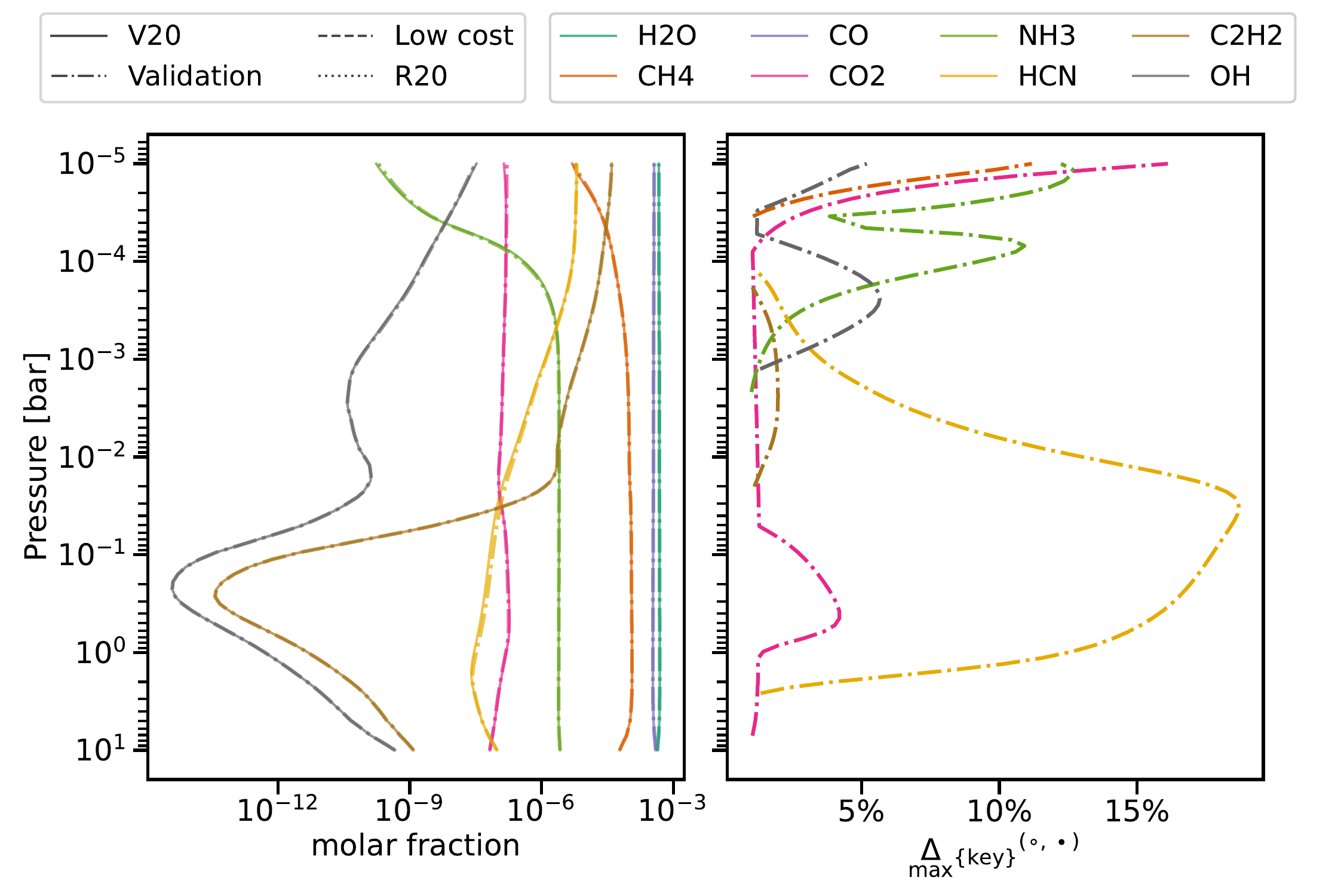}
        \caption{}
    \end{subfigure}
    
    \caption{Same as Fig. ~\ref{fig:key} but for models in the presence of photochemistry, for HD 209458b (a) and HD 189733b (b).}
    \label{fig:photo}
\end{figure*}

\subsection{Photoschemes}

Our final and most challenging test involved incorporating photochemistry into the model reduction process, and it yielded highly promising results. Although DARWEN does not currently alter photochemical reactions directly, we applied photochemistry and allowed the 1D model to determine whether to include a photochemical reaction based on the presence of the involved species. The full model with photochemistry consists of 1906 ``normal'' reactions plus 52 photochemical reactions, totaling 1958 reactions (as shown in Table~\ref{tab:losses}). The best scheme obtained by DARWEN, shown in Fig.~\ref{fig:photo}, includes 48 molecules and 722 ``normal'' reactions (37 $\%$ of the total number of normal reactions), and 34 photochemical reactions corresponding to the species present in the chemical scheme (65 $\%$ of the total number of photochemical reactions).

As explained earlier, we ran DARWEN to optimize the test planets' atmospheres simultaneously. The resulting scheme attained accuracies of 0.16 and 0.18 for key molecules, with computation times of just 33 and 43 seconds for HD 209458b and HD 189733b, respectively. These computation times represent a substantial improvement over the full model, highlighting the practical advantages of our reduced schemes. Interestingly, $\mathrm{HCN}$ emerged as a major difference molecule again, alongside $\mathrm{CH}_4$ and $\mathrm{CO}_2$. The latter two molecules exhibited significant differences in the upper atmosphere, indicating a strong sensitivity to photochemistry, while $\mathrm{HCN}$ showed minor discrepancies in the lower layers, consistent with our previous low-cost schemes.

While the accuracy of the photoscheme was not as high as that of the validation scheme — being roughly two to three times lower — the abundances remain qualitatively well reproduced, as illustrated in Fig. ~\ref{fig:photo}. The discrepancies observed are still within acceptable ranges, especially considering that they may be smaller than current uncertainties arising from factors such as temperature, metallicity, and mixing rates, as well as uncertainties from unknown reactions, Arrhenius expression fitting of rate constants, and the eddy diffusion parameterization. This suggests that our reduced model with photochemistry is robust and reliable for practical applications.

Integrating photochemistry into the reduction process and achieving a consistent reduced scheme marks a significant advancement in this field, demonstrating DARWEN's ability to handle the complexities introduced by photochemical reactions. While the promising results indicate effectiveness, the observed lower accuracy highlights areas for further refinement, such as identifying and including critical missing reactions. Our data-driven approach is designed for iterative improvement, and future enhancements, including the incorporation of intrinsic model uncertainties, could lead to more reliable decision-making and models. Overall, this integration showcases DARWEN's strong potential, paving the way for more accurate and efficient chemical modeling of exoplanetary atmospheres.

\section{Conclusions}\label{conclusions}

We have developed DARWEN, a GA designed to efficiently reduce complex chemical networks for exoplanetary atmospheres. This method could potentially be applied to other atmospheric chemical contexts, such as those found in Solar System planets. We have demonstrated its ability to achieve accuracies similar to previous reduced models while significantly enhancing computational efficiency.

We applied DARWEN to reduce the full chemical model from \citet{Venot2020}, V20, and compared our results to their associated reduced model, R20. We generated three types of reduced schemes: a validation scheme, a low-cost scheme, and a photoscheme, each with a different aim. The validation scheme focused on demonstrating DARWEN's capacity to enhance the accuracy of key molecules. The low-cost scheme aimed to improve computational efficiency by relaxing precision requirements. The photoscheme was designed to be the first reduced scheme in the field that includes photochemical reactions.

The validation scheme has a precision comparable to that of R20, with superior accuracy for the major species (maximum discrepancy of 58 $\%$ compared to 98 $\%$ for R20). The low-cost scheme used the validation scheme as the initial solution and focused on minimizing the number of molecules while maintaining similar accuracy levels. This scheme ran 2.5 times faster than R20 and 10 times faster than V20, demonstrating DARWEN's capability to create more efficient schemes. Although some discrepancies were observed in $\mathrm{HCN}$ (with a maximum $\Delta$ of 33$\%$), this is still within our acceptable uncertainty constraints. Our objective of obtaining a fast new scheme for a particular set of planets was achieved.

Our most challenging task was using DARWEN to develop the photoscheme, which involved incorporating photochemistry into the model reduction process. Although we achieved lower accuracies compared to our validation scheme and R20 — which cannot be applied when photochemistry is involved — the photoscheme represents a pioneering step in exoplanet research because it is the first time a chemical network that includes photochemical reactions has been reduced. The resulting schemes reached discrepancies of up to 18 $\%$ in $\underset{\text{max}}{\Delta}$ for key molecules, with computation times of 33 and 43 seconds for HD 209458b and HD 189733b, respectively — more than 20 times faster than the full model. The challenges in optimizing the scheme for multiple planets, especially when incorporating photochemistry, highlight areas for further refinement.

Additionally, incorporating intrinsic model uncertainties into the decision-making process for evaluating discrepancies between the full and reduced models could enhance our approach. Considering that all full chemical schemes possess inherent limitations regarding the accuracy of their rate constants and other parameters derived from the literature, understanding the intrinsic uncertainties of the model is essential to assessing whether a discrepancy of 18$\%$ (or 33$\%$ in the case of the low-cost scheme) is significant. Nevertheless, we are optimistic about these results, as they mark a significant advancement in incorporating photochemistry into reduced chemical networks at a time when photochemical processes are increasingly important in the characterization of exoplanets.

While DARWEN has proven effective in reducing chemical networks for the specific exoplanets we have studied, applying it to other planets currently requires conducting a new run of the algorithm with the necessary planetary data for each case. This means that for detailed models — such as those needed for 3D atmospheric simulations — we must input the relevant data into DARWEN to generate a new, tailored chemical scheme for each planet or group of planets of interest. In future work, we plan to expand this approach by applying DARWEN to a selection of exoplanets observed or expected to be observed with JWST and other facilities. By examining how the resulting chemistry of these planets varies, we aim to gain deeper insight into the diversity of exoplanetary atmospheres and, hopefully, construct a more comprehensive chemical scheme that can be applied more broadly.

Overall, DARWEN is a powerful tool for optimizing chemical network reductions, effectively balancing precision and computational efficiency. Its application to exoplanetary atmospheres opens new avenues for exploring and understanding the complex chemistry of these distant worlds. By developing these reduced networks, we enable the inclusion of detailed kinetic networks in complex models such as GCMs, thereby enhancing our understanding of exoplanets.

\begin{acknowledgements}
 OV and RV acknowledge funding from Agence Nationale de la Recherche (ANR), project ``EXACT'' (ANR-21-CE49-0008-01). In addition, OV acknowledges funding from the Centre National d’Études Spatiales (CNES). ALB, JNH, TK, RB and LD acknowledge KULeuven funding of the ID-N ESCHER project. 
\end{acknowledgements}

%
%

\bibliographystyle{aa} 
\bibliography{references}

\appendix

\section{Sensitivity analysis and rate constant variations}\label{app:reacs}

\begin{figure*}[ht]
   \centering

   \includegraphics[width=\textwidth]{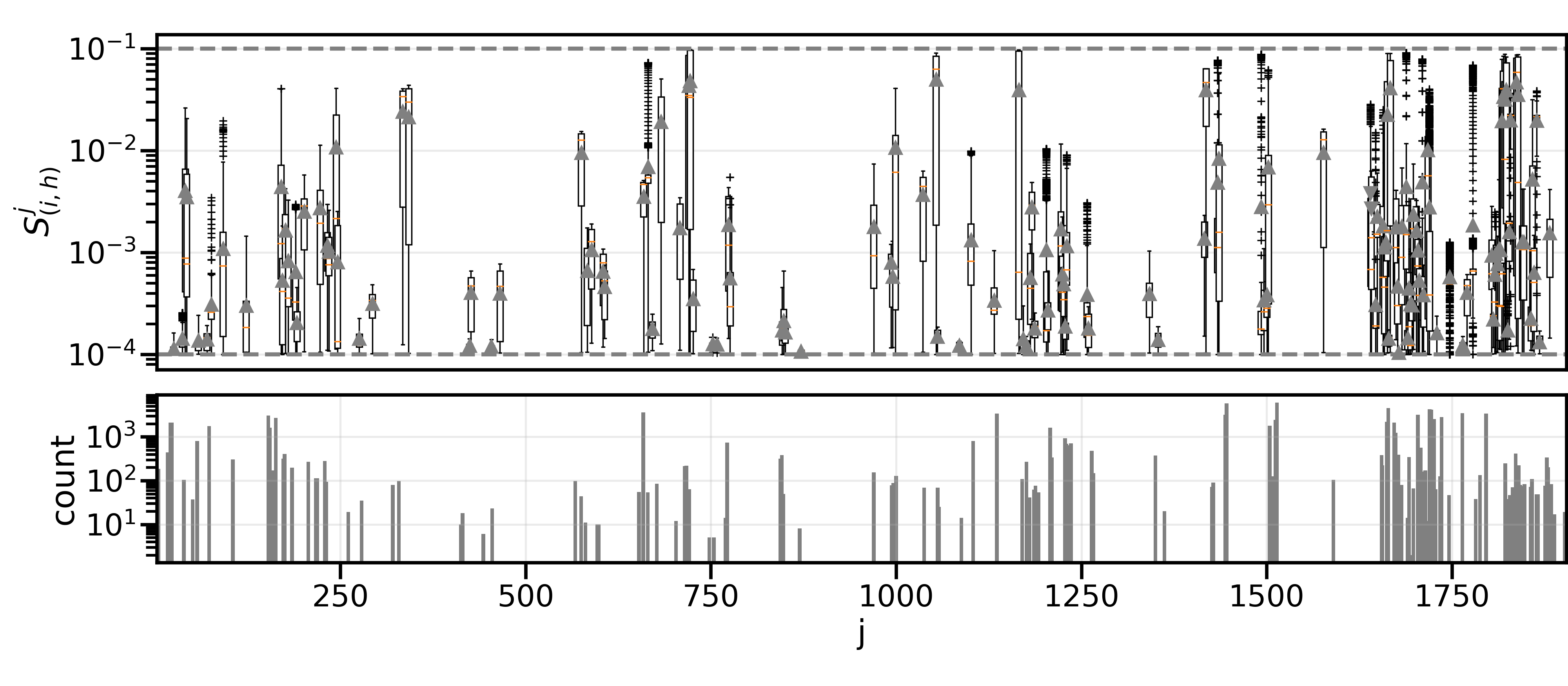}
   \caption{{Sensitivity analysis of HD 209458b without photochemistry:} \textit{Top}: Distribution of sensitivities for all molecules at all heights for each reaction ID ($j$), showing only those above the expected ODE solver noise level. Horizontal lines indicate the noise level and 10 $\%$ variation, the maximum expected if the response is linear. \textit{Bottom}: Count of sensitivities above the noise level for each reaction.}

              \label{fig:sensis}
\end{figure*}

    The code receives coefficients to build the rate constants according to the modified Arrhenius equation: 

\begin{equation} \label{eq:almostarrhenius}
    k_f(T) = \alpha T^{\beta} \exp \left( \frac{-\gamma}{T} \right)
,\end{equation}where $k_f$ is the forward rate constant, $T$ is the temperature, $\gamma$ is the activation energy divided by the Boltzmann constant, and $\alpha$ and $\beta$ are coefficients. To obtain the reverse reaction rate constant $k_r$ of a given reaction, we used the fact that the constant of equilibrium, $K_{eq}$, is related to the standard Gibbs energy, $\Delta G^0$, and standard pressure, $P^0$,
\begin{equation}
    K_{eq} = \frac{k_f}{k_r} = \exp \left( \frac{-\Delta G^0}{RT}\right) \left( \frac{P^0}{k_BT} \right)^{\Delta \nu}
,\end{equation}where $k_B$ is the Boltzmann constant and R is the Rydberg constant.

To assess the sensitivity of reactions that can be modeled as Eq.~\ref{eq:almostarrhenius} we varied their alpha coefficient, $\alpha$, such that $\alpha^*= 1.1 \alpha $. $\beta$ and $\gamma$ are unchanged. The choice of $\Delta \alpha = 0.1 \alpha$ strikes a balance between ensuring the alteration is sufficiently small to approximate a derivative, yet large enough to noticeably affect species abundances given the  achievable steady-state convergence threshold. Also, to maintain the thermodynamics unvaried, for each forward reaction varied we varied the corresponding reverse reaction rate constant parameter by the same factor. We note that for some reactions, the rate constant as a function of temperature is expressed in a different way \citep[see][]{Venot2012}. For all reactions, we used the same multiplicative perturbation approach so that $k_f$ is multiplied by 1.1 at each temperature.

A previous study advises against using variations of more than 5$\%$ for ``local sensitivity analysis'' \citep{Saltelli2005, Dobrijevic2010} for related sensitivity analyses, but we have found only modest changes in terms of predicted model reduction properties between our chosen 10$\%$ variations and a smaller 5$\%$ change, with the molecules and reactions selected in the PCA-based reduced scheme being only slightly changed. Further changes following the GA component of the algorithm will still further reduce these differences.

\section{Genetic algorithm details}\label{app:GA}

The iterative process at each generation involves the following steps:

\begin{enumerate}
    \item \textit{Selection:} At each generation, $n_{fit}$ schemes are chosen as progenitors. This can be done deterministically by selecting the $n_{fit}$ fittest schemes or using a {weighted roulette rule}, where selection probability is $\varphi_s / \sum_s \varphi_s$. In the first iteration, $n_{fit}=1$.
    
    \item \textit{Crossover:} Each progenitor has a probability $p_c$ of crossover. We applied a one-point crossover by selecting a random position in the binary gene sequence and swapping genes beyond this point with another progenitor.

    \item \textit{Self-mutation:}  Each progenitor has a mutation probability $p_m$. About $p_m \times n_{pop}$ schemes mutate per generation, flipping a random number of genes (up to $p_m \times N_{reacs}$). Mutations may introduce new reactions, adding the necessary molecules to the species list, increasing diversity.

    \item \textit{Exploring the species pool:} Each progenitor generates $n_{child}$ new schemes. Reaction inclusion is guided by mutation rate $\rho_m$ and restricted reaction pool, with the killing rate $\rho_{\dagger}$ controlling how many reactions are excluded. Our code sets species number limits, giving each new individual an equal chance to fall within this range. Molecules are randomly removed if fewer species are generated, or added if more are needed. Key species are always retained to balance computational cost and accuracy.

    \item  \textit{Performance evaluation:} All individuals are evaluated according to Eq.~\ref{eq:loss}.

    \item  \textit{Elitism:} Every $n_{elit}$ cycles, if performance drops, the $n_{fit}$ fittest individuals from previous generations are selected, preserving valuable information.

    \item \textit{Curiosity mechanism:} To explore neglected regions of the solution space, we rewarded the inclusion of under-explored reactions (i.e., those added fewer than $N_{cur}$ times to fit schemes). Optionally, $n_{cur}$ such reactions can be added to the initial scheme, increasing variability and promoting thorough exploration. This optional step, original to our work, helps increase the variability of schemata and ensures the exploration of neglected solution spaces.

\end{enumerate}

\begin{table*}[t]
\centering
\caption{Glossary of terms related to our GA.}
\label{tab:glossary}
\begin{tabular}{lll}
\hline\hline

\textbf{Term} & \textbf{Value} & \textbf{Definition} \\

\hline
$N_{reacs}$& 958 & Total number of {forward}  reactions in our tested network \\
$n_{reacs}$& $\sim$ 150 - 600 & Total number of forward and reverse reactions tested per scheme \\
$n_{fit}$& 15-20 & Number of progenitors selected per generation \\
$n_{child}$& 15-20 & Number of children generated from each progenitor \\
$n_{pop}$& $n_{fit} \times n_{child}$ & Total number of individuals per  generation \\
$p_c$& 0.5 & Probability of crossover \\
$p_m$& 0.05 & Probability of self-mutation \\

$\rho_m$ & $\sim$0.15 & Mutation rate, determining how many reactions from the pool are added. \\
$\rho_{\dagger}$ & $\sim$ 0.1 & Killing rate, determining how many reactions from the individual are excluded. \\
$n_{mlc}$& $\sim$ 35-55 & Number of molecules per scheme \\
$n_{elit}$& $\sim$ 7 & Frequency of elitism application (generations) \\
$n_{cur}$& $\lesssim$ 5 & Number of curious reactions added per scheme \\
$N_{cur}$& 100 & Upper threshold for defining an under-explored reaction \\ \hline
\end{tabular}

\end{table*}

\section{Major species discrepancies}\label{app:major}
\begin{figure*}[ht]
    \centering
    
    \begin{subfigure}[b]{0.49\textwidth}
        \centering
        \includegraphics[width=\textwidth]{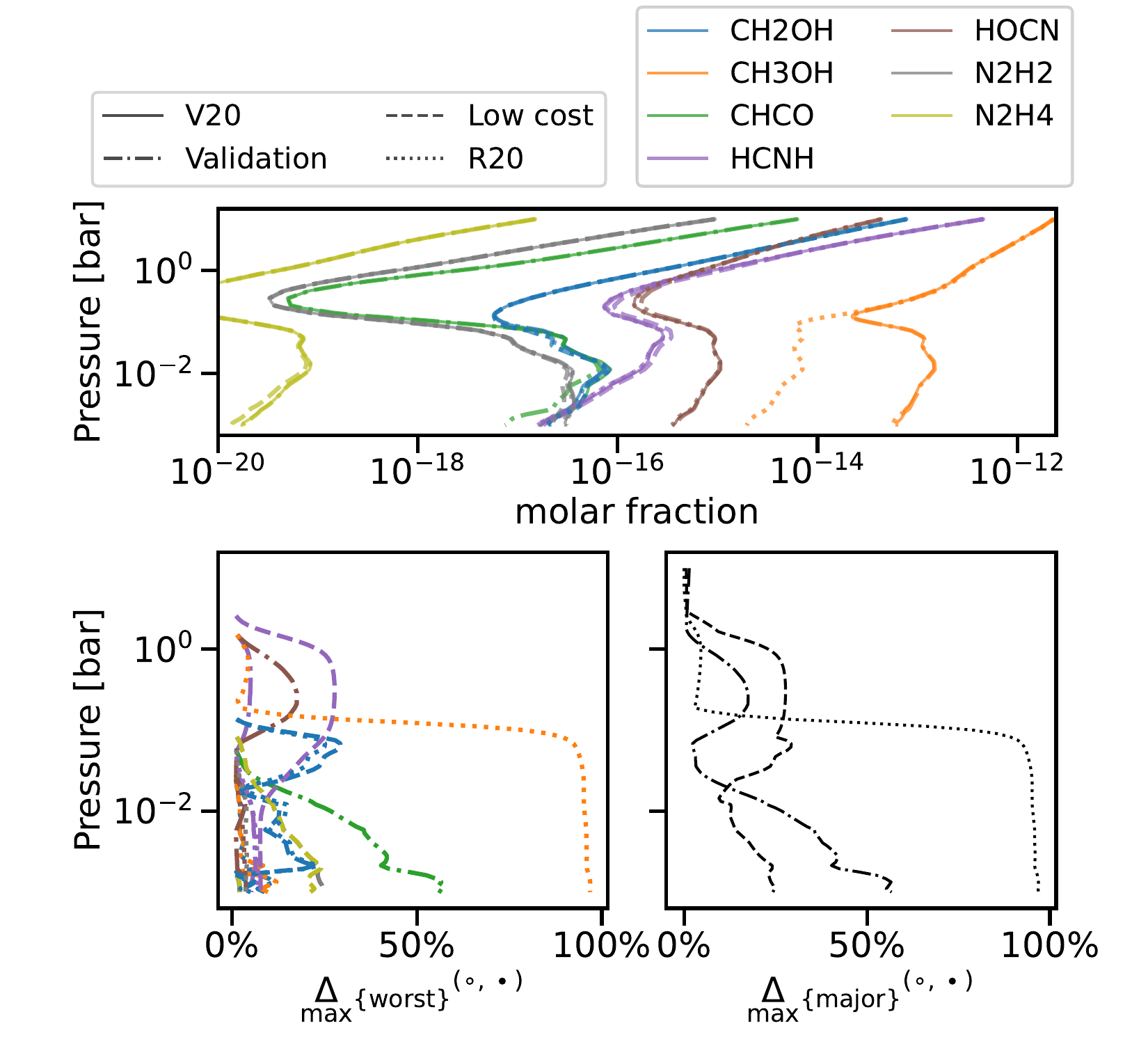}
        \caption{}
    \end{subfigure}
    \hfill
    \begin{subfigure}[b]{0.49\textwidth}
        \centering
        \includegraphics[width=\textwidth]{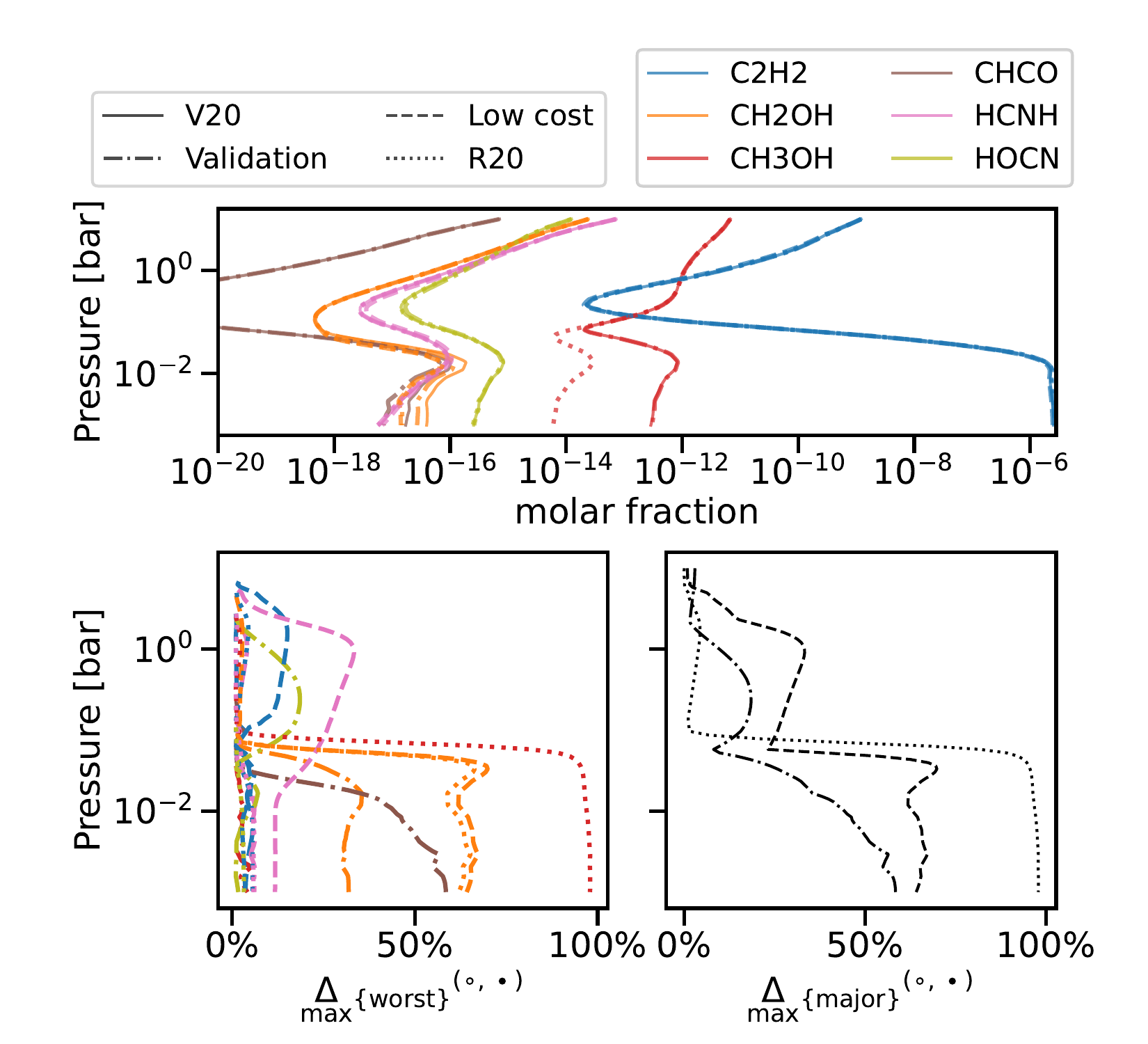}
        \caption{}
    \end{subfigure}
    
    \begin{subfigure}[b]{0.49\textwidth}
        \centering
        \includegraphics[width=\textwidth]{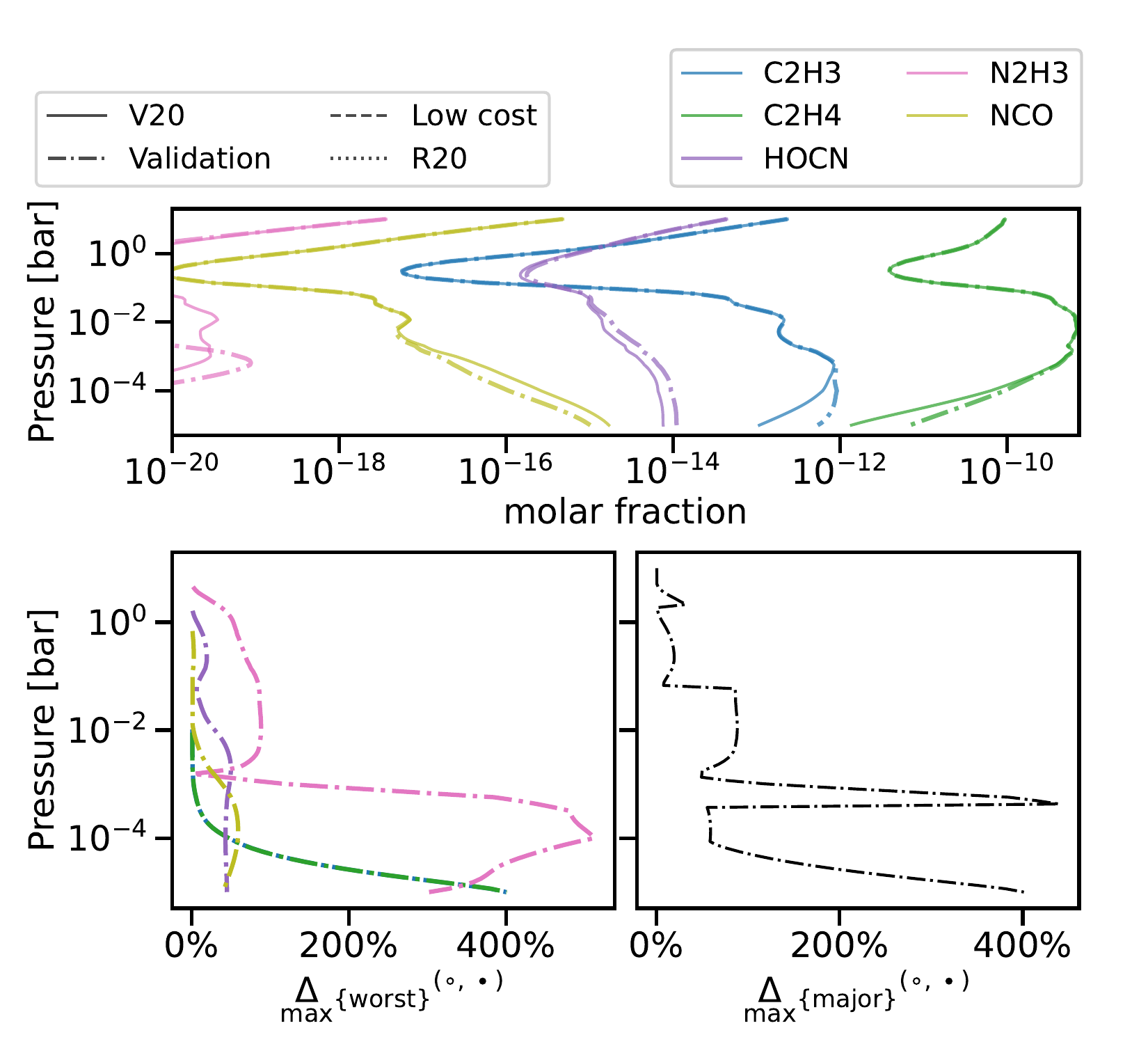}
        \caption{}
    \end{subfigure}
    \hfill
    \begin{subfigure}[b]{0.49\textwidth}
        \centering
        \includegraphics[width=\textwidth]{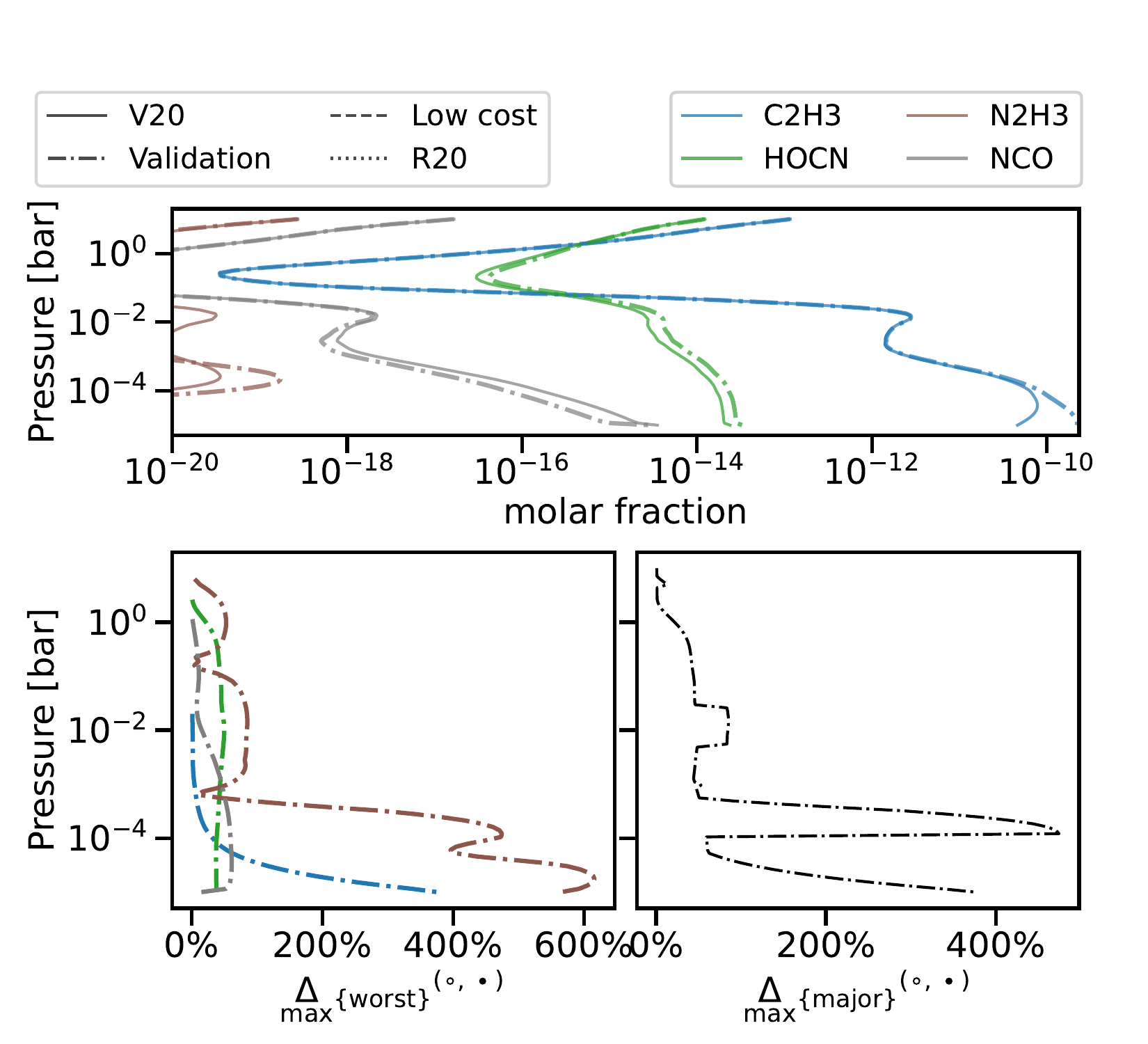}
        \caption{}
    \end{subfigure}
    
    \caption{{Discrepancies on selected major species}. Major species with discrepancies of more than 10$\%$ compared to the full model at any height are shown for (a) HD 209458b (without photochemistry),  (b) HD 189733b (without photochemistry), (c) HD 209458b (with photochemistry), and (d) HD 189733b (with photochemistry). \textit{Top panels}: Absolute abundances. \textit{Bottom left}: Vertical profile of $\underset{\text{max}}{\Delta}$ for each subset of species. \textit{Bottom right}: Maximum discrepancy for those species at each height. Note that in HD 189733b with photochemistry (d), the maximum discrepancy does not always map the species present because $\mathrm{N}_2 \mathrm{H}_3$ is defined as a major species only at certain pressure levels.}
    \label{fig:worst}
\end{figure*}

In the main text, we mention that our criterion for choosing between different promising schemes identified by our GA-based approach relates to major species discrepancies. Here, we present the species that exhibit a discrepancy of over 10$\%$ at least at one height of the model, as shown in Fig. \ref{fig:worst}. This criterion for our final decision is optional; we could have chosen the scheme with the smallest $\underset{\text{max}}{\Delta}$. However, $\underset{\text{max}}{\Delta}$ did not seem to correlate well with major species abundances beyond a certain extent. For example, Fig.~\ref{fig:worst} shows that some major species in R20 have discrepancies of almost 100$\%$. The reasons why these discrepancies do not correlate and how they compare to the uncertainties in the model require follow-up studies.

\end{document}